\documentstyle[12pt]{article}
\begin{document}

\title {\bf Complexity in some Physical Systems}
\author{R. L\'{o}pez-Ruiz \\
                                   \\
\small Dpto. de F\'{\i}sica Te\'{o}rica,\\
\small Facultad de Ciencias, Universidad de Zaragoza,\\
\small 50009-Zaragoza, Spain}
\date{ }

\maketitle
\baselineskip 8mm

\begin{center} {\bf Abstract} \end{center}
The LMC-{\it complexity} introduced by L\'{o}pez-Ruiz,
Mancini and Calbet [Phys. Lett. A {\bf 209}, 321-326 (1995)]
is calculated for different
physical situations: one instance of classical statistical mechanics,
normal and exponential distributions, and a simplified laser model.
We stand out the specific value of the population inversion
for which the laser presents maximun complexity.

\newpage

\section{Introduction}

The notion of {\it complexity} in physics comes from considering
the perfect crystal and the isolated ideal gas as examples of
simple models, and therefore as systems with zero complexity. The
asymptotic properties that an indicator of complexity should have
are induced from the characteristics of the two former systems. We
have proposed in \cite{lopez,calbet} a simple mathematical
expression to calculate this quantity. The result was to define
{\it LMC-complexity} ($C$) as the interplay between the {\it
information} ($H$) stored in a system and its distance to the
equipartition, called {\it disequilibrium} ($D$):
\begin{equation}
C = H \cdot D = -\left ( K\sum_{i=1}^{M} p_i\log p_i \right ) \cdot
\left (\sum_{i=1}^{M}(p_i - \frac{1}{M})^2 \right ),
\end{equation}
where $\{p_1,p_2,...,p_M\}$ represent the probabilities of the $M$
states $\{x_1,x_2,...,x_M\}$ accessible to the system, with the
normalization condition $\sum_{i=1}^{N}p_i=1$; $H$ gives the Shannon
information, where $K$ is a constant \cite{shannon} and
$D$ is a measure of a probabilistic hierarchy
among the states. This definition presents advantages since
it is based upon a probabilistic description of the system,
and the knowledge of the probability distribution
of its accesible states allows to calculate the LMC-complexity.
The quantity $C$ is associated to a particular description and therefore
to a scale of observation, just as intuition suggests.

In the following sections we calculate $C$ in different
physical situations. Sections 2. and 3. are devoted to obtain
LMC-complexity for a system in thermal equilibrium and for
the normal and the exponential probability distributions.
In Section 4. the previous results are applied to study the behavior
of $C$ in a simplified model of a two-level laser system.

\section{Complexity in the Canonical Ensemble}

Each physical situation is closely related to a specific
distribution of microscopic states. Thus, an isolated
system presents equipartition, by hypothesis: the microstates
compatible with a macroscopic situation are equiprobable \cite{huang}.
The system is said to be in equilibrium.
For a system surrounded by a heat reservoir the probability of
the microstates associated to the thermal equilibrium follow the
Boltzmann distribution.
In general, the stablished scheme consists in associating a
probability distribution of states to each phenomenon.
If the system presents
some specific distribution it is said to be
in some kind of {\it equilibrium}.
From this point of view, complexity $C$ can be assigned to each
system depending on the specific description.

Let us try to analyse the behavior of $C$ in an ideal gas
in thermal equilibrium. In this case the probability $p_i$
of each accesible state is given by the Boltzmann distribution:
\begin{eqnarray}
p_i & = & \frac{e^{-\beta E_i}}{Q_N}, \\
Q_N & = & \int e^{-\beta E(p,q)} \frac{d^{3N}p d^{3N}q}{N! h^{3N}} =
e^{-\beta A(V,T)},
\end{eqnarray}
where $Q_N$ is the partition function of the canonical ensemble,
$\beta=1/\kappa T$ with $\kappa$ the Boltzmann constant and $T$ the
temperature, $V$ the volume, N the number of particles,
$E(p,q)$ the hamiltonian of the system, $h$ is the Planck constant
and $A(V,T)$ the Helmholtz potential.

Calculation of $H$ and $D$ gives us:
\begin{eqnarray}
H(V,T) & = & ( 1+T\frac{\partial}{\partial T})
\left( \kappa\log Q_N \right)  =   S(V,T), \\
D(V,T) & = & e^{2\beta \; \left[ A(V,T) - A(V,T/2) \right]}.
\end{eqnarray}
Note that Shannon information $H$ coincides with the thermodynamic
entropy $S$ when $K$ is identified with $\kappa$. If a system
verifies the relation $U=C_v T$ ($U$ the internal energy, $C_v$ the
specific heat) the complexity takes the form:
\begin{equation}
C(V,T) \sim cte(V) \cdot S(V,T) e^{-S(V,T)/\kappa}
\end{equation}
that matches the intuitive function proposed in Figure 1.

\section{Complexity in Distributions}

The introduced indicator, LMC-complexity,
is closely related to the probability distribution of states
associated to a system. We remark this idea calculating
$C$ for the normal and exponential
distributions.

{\bf Normal Distribution}: Suppose a continuum of states represented
by the $x$ variable whose probability density $p(x)$ is given by the
normal distribution of variance $\sigma$:
\begin{equation}
p(x)=\frac{1}{\sigma \sqrt{2\pi}}\exp\left(-\frac{x^2}{2\sigma^2}\right).
\end{equation}
The expressions obtained for $H$, $D$ and $C$ are the following: \par
\begin{eqnarray}
H & = & -K\int_{-\infty }^{+ \infty }p(x)\log p(x) \;\; = \;\;
K \left(\frac{1}{2} + \log (\sigma \sqrt{2\pi })\right),   \nonumber \\
D & = & \int_{-\infty}^{+ \infty}p^2(x)dx,  \;\; = \;\;
\frac{1}{2\sigma \sqrt{\pi}} \nonumber \\
C_g & = & H \cdot D \;\; = \;\; \frac {K}{2\sigma\sqrt{\pi}}\left(
\frac {1}{2}+\log(\sigma \sqrt{2\pi})\right).
\end{eqnarray}

The additional condition $H\geq 0$ imposes
$\sigma\geq\sigma_{min}=(2\pi e)^{-1/2}$. The highest complexity is reached
for a determined width: $\bar{\sigma}=\sqrt{(e/2\pi)}$.

{\bf Exponencial Distribution}: Consider a exponencial
distribution of variance $\gamma$:
\begin{equation}
p(x)=\left\{\begin{array}{ll}
\frac{1}{\gamma}e^{-x/\gamma} & x>0 \\
0                             & x<0.
\end{array} \right.
\end{equation}

The same calculation gives us:
\begin{eqnarray}
H & = & K (1 + \log\gamma), \nonumber  \\
D & = & 1/2\gamma, \nonumber \\
C_e & = & \frac{K}{2\gamma}(1+\log\gamma).
\end{eqnarray}
with the condition $\gamma\geq\gamma_{min}=e^{-1}$. The highest complexity
corresponds in this case to $\bar{\gamma}=1$.

In Fig. 2. the dependence of $C$ on width ($\sigma=\gamma$) is
represented. Remark that for the same width the exponential
distribution presents a higher complexity ($C_e/C_g \sim 1.4$).

\section{Complexity in a Two-Level Laser Model}

One step further, combining the results obtained in Secs. 2. and 3.,
is now done. We calculate LMC-complexity for an unrealistic and
simplified model of laser \cite{svelto}.

Let us suppose a laser of two levels of energy: $E_1=0$ and $E_2=\epsilon$,
with $N_1$ atoms in the first level and $N_2$ atoms in the second level,
and the condition $N_1+N_2=N$ (the total number of atoms) (Fig. 3.).
Our aim is to sketch the statistics of this model and
to introduce the results of photon counting \cite{arecchi}
that produces an asymmetric behavior of $C$ as function of
the population inversion $\eta =N_2/N$. In the range
$\eta\in (0,1/2)$ spontaneous and stimulated emission can take place,
but only in the range $\eta\in (1/2,1)$ the condition to have
lasing action is reached, because the population must be, at least,
inverted, $\eta>1/2$.

The entropy ($S$) of this system vanishes when $N_1$ or $N_2$ is zero.
Moreover, $S$ must be homegenous of first order in the extensive
variable $N$ \cite{callen}.
For the sake of simplicity we approach $S$ by the first term in the
Taylor expansion:
\begin{equation}
S\sim \kappa \frac{N_1N_2}{N}= \kappa N\eta (1-\eta).
\end{equation}
The internal energy is $U=N_2\epsilon =\epsilon N\eta$ and the statistical
temperature is:
\begin{equation}
T=\left(\frac{\partial S}{\partial U}\right)_N^{-1}=
\frac{\epsilon}{\kappa}\frac{1}{(1-2\eta)}.
\end{equation}
Note that for $\eta>1/2$ the temperature is negative as corresponds
to the stimulated emission regime dominating the actual laser action.\newline
From Eq. (5) the value of disequilibrium is in this case:
\begin{equation}
D(N,T)=e^{N/4}e^{-2S/\kappa},
\end{equation}
and then LMC-complexity is:
\begin{equation}
C= e^{N/4}\cdot Se^{-2S/\kappa}.
\end{equation}
In the laser regime, the quantity $C$ can reach the same order
of magnitude of $S$ when $D\sim 1$.

We are now interested in introducing qualitatively
the results of laser photon counting in the calculation of
LMC-complexity. It was reported in \cite{arecchi} that the
photo-electron distribution of laser field appears to be poissonian.
In the continuous limit the Poisson distribution is
approached by the normal distribution \cite{distri}.
The width ($\sigma$) of this energy
distribution in the canonical ensemble is proportional to the statistical
temperature of the system. Thus, for a switched on laser in the regime
$\eta\in [1/2,1]$, the width of the gaussian energy distribution
can be fitted by choosing $\sigma\sim -T \sim 1/(2\eta-1)$
(recall that $T<0$ in this case).
The range of variation of $\sigma$ is
$[\sigma_{\infty},\sigma_{min}]=[\infty, (2\pi e)^{-1/2}]$.
Then we obtain:
\begin{equation}
\sigma \sim\frac{(2\pi e)^{-1/2}}{2\eta-1}.
\end{equation}
By replacing this expression in Eq. (8),
and rescaling by a factor proportional
to entropy, $S\sim\kappa N$, (in order to give to it the correct
order of magnitude), LMC-complexity for a population inversion
in the range $\eta\in [1/2,1]$ is reobtained:
\begin{equation}
C_{laser}\simeq \kappa N \cdot (1-2\eta)\log(2\eta-1).
\end{equation}
We consider at this level of discussion
$C_{laser}=0$ for $\eta<1/2$.
The behavior of this function for the whole range of parameter
$\eta\in [0,1]$, is plotted in Fig. 4.
It is worth noticing the value
$\eta_2 \simeq 0.68$ where the laser presents
the highest complexity.

By following theses ideas, if the width, $\sigma$, of the
experimental photo-electron distribution of laser field
is measured, the population inversion parameter, $\eta$,
would be given by Eq. (15). In a next step, we would obtain
the LMC-complexity of the laser system by Eq. (16).

\section{Conclusions}

A model helps us to approach the reality and provides invaluable
guidance in the objetive of a finer understanding of a physical
phenomenon. From this point of view the present work tries to
enlighten the problem of calculating the {\it LMC-complexity},
$C$, of a physical system via a simplified model. Thus, an
unrealistic presentation of a two-level laser system has been
worked out. In this context, we have obtained an expression for
the quantity $C$ as a function of the population inversion,
$\eta\in [0,1]$. The laser presents the highest complexity for
$\eta\simeq 0.68$. A formal experimental approach to its
measurement, if possible, is proposed.

{\bf Acknowledgements}
The author wants to thank Prof. H.L. Mancini, from Universidad
Privada de Navarra (Spain), for very useful and fruitful discussions.

\newpage
\begin{center} \bf Figure Captions \end{center}

{\bf Fig 1.} Sketch of the intuitive notion for the magnitudes:
{\it information} (H), {\it disequilibrium} (D) and {\it complexity }
for the physical systems.
Extreme systems are the perfect crystal and the isolated ideal gas. \par

{\bf Fig 2.} LMC-Complexity ($C=H\cdot D$) as a function of the width
($\sigma = \gamma$) for the normal and exponential distributions.
(It is suppossed $K=1$). \par

{\bf Fig 3.} Diagram showing the two-level laser model used in Sec. 4. \par

{\bf Fig 4.} LMC-Complexity on the population inversion ($\eta$) for the two-level
laser model.  Observe the peak in complexity for
$\eta_2 \simeq 0.68$ (lasing regime).
(Units in $C$ scaled by $\kappa N$).

\end{document}